\documentclass[twocolumn,prb,showpacs,superscriptaddress,amsmath,amssymb,natbib]{revtex4}

\usepackage[dvips]{epsfig}
\usepackage{graphicx}
\usepackage{dcolumn}
\usepackage{bm}
\usepackage[english,german]{babel}
\usepackage{amsmath}
\usepackage{amssymb}
\usepackage[latin1]{inputenc}

\newcommand{\mueof}{\mu_{_{\mbox{\tiny EOF}}}}
\newcommand{\veof}{v_{_{\mbox{\tiny EOF}}}}
\newcommand{\vveof}{{\bf v}_{_{\mbox{\tiny EOF}}}}

\newcommand{\gDPD}{\gamma_{{\mbox{\tiny DPD}}}}
\newcommand{\oDPD}{\omega_{{\mbox{\tiny DPD}}}}

\begin{document}

\title{Polyelectrolyte Electrophoresis in Nanochannels: A Dissipative Particle
  Dynamics Simulation}

\author{Jens Smiatek}
\email{jens.smiatek@uni-muenster.de}
\affiliation{Institut f{\"u}r Physikalische Chemie, Universit{\"a}t
    M{\"u}nster, D-48149 M{\"u}nster, Germany} 

\author{Friederike Schmid}
\email{Friederike.Schmid@Uni-Mainz.DE}
\affiliation{\selectlanguage{german}
  Institut f"ur Physik, Johannes Gutenberg-Universit"at, Staudinger Weg 7, D-55099
  Mainz, Germany
}

\selectlanguage{english}

\date{\today}

\begin{abstract} 
We present mesoscopic DPD-simulations of polyelectrolyte electrophoresis in
confined nanogeometries, for varying salt concentration and surface slip
conditions. Special attention is given to the influence of electroosmotic 
flow (EOF) on the migration of the polyelectrolyte. The effective polyelectrolyte 
mobility is found to depend strongly on the boundary properties, {\em i.e.}, 
the slip length and the width of the electric double layer. Analytic expressions 
for the electroosmotic mobility and the total mobility are derived which are 
in good agreement with the numerical results. The relevant quantity 
characterizing the effect of slippage is found to be the dimensionless 
quantity $\kappa \: \delta_B$, where $\delta_B$ is the slip length, 
and $\kappa^{-1}$ an effective electrostatic screening length at the channel 
boundaries. 
\end{abstract}



\maketitle

\section{Introduction}
\label{intro}

In recent years there is growing interest in techniques for manipulating single
nanoparticles or macromolecules in micro- and nanochannel systems. The flow
profiles in these channels and the motion of the macromolecules can be
controlled on the nanoscale by pressure gradients and electric fields, and by
exploiting smart channel geometries. This explains the great potential and the
broad applicability of nanochannel devices, {\em e.g.}, for analyzing tiny DNA
or protein samples by electrophoresis 
\cite{Viovy00,Iki96,Roer97,Mathe07,effenhauser97,bader99,han00,han02,huang02,duong03,ros04}.
 
Such systems represent a challenge for theory and computer simulation due to
their high complexity. In biotechnological applications, the molecules of
interest are often charged and dissolved in buffer solutions with high salt
concentrations. Thus electrostatic and hydrodynamic effects compete with each
other, resulting, among other, in a remarkable 'electrohydrodynamic screening'
effect: In free solution electrophoresis ({\em i.e.}, in a constant electric
field), the counterion layer surrounding a charged particle not only screens
the electrostatic interactions, but also the dominant contribution to the
hydrodynamic interactions, {\em i.e.}, those that are generated by the applied
field \cite{Manning81,Barrat96}.  Therefore, simulations of electrophoresis
that neglect the electrostatic and hydrodynamic interactions altogether often
give results that are in good semiquantitative agreement with experiments
\cite{Viovy00,SSDR04,SSDR05}. More sophisticated approaches that still allow to
avoid the explicit representation of charges have been devised as well
\cite{Duong-Hong08,Slater09}. Nevertheless, it is clear that such simplified
treatments disregard important physics, especially in confined geometries where
electrostatic interactions compete with the regular steric interactions with the
confining walls \cite{Long96}. Simulations that take full account of
electrostatic and hydrodynamic interactions are clearly desirable. Such
simulations have recently been carried out for polyelectrolyte electrophoresis
in free solutions \cite{GBS08,Frank08}, but, to the best knowledge of the present
authors, not yet for microchannels.

In many microchannels, an additional effect comes into play, which
significantly modifies the effective electrophoretic response of particles to
electric fields: The same electric field that drives the polyelectrolyte may
also induce a total net fluid flow in the microchannel, the 'electroosmotic
flow' (EOF).  Many materials commonly used in microtechnology like PDMS
(Polydimethylsiloxane) acquire charges if brought in contact with water, either
by the ionization or dissociation of surface groups or the adsorption of ions
from solution. To screen the charges on the channel walls, a diffuse layer of
oppositely charged ions forms in front of the walls. In the presence of an
external electric field ${\bf E}^{(ext)}$, these ions are pulled along,
dragging the surrounding fluid with them. One gets a characteristic 'plug' flow
profile which saturates at a fluid velocity 
\begin{equation} \label{eq:mueof}
\vveof = \mueof {\bf E}^{(ext)} 
\end{equation} 
outside of the diffuse layer, with the so-called electroosmotic mobility
$\mueof$.  The effective migration speed ${\bf v}_P$ of particles in
microchannels in response to the external fields then results from two
contributions \cite{SSDR05}: The bare electrophoretic mobility $\mu_e$
of the polyelectrolyte in a fluid at rest, and the background EOF velocity,
${\bf v}_P = \vveof + \mu_e {\bf E}^{(ext)}$.  Experimentally, it has been
found that the former may even dominate over the latter, such that the
polyelectrolyte effectively migrates in the direction {\em opposite} to the
applied field \cite{Mathe07}.

Now if the diffuse layer is thin compared to the channel dimensions,
Eq.~(\ref{eq:mueof}) can be regarded as an effective boundary condition for a
steady-state EOF velocity field $\vveof({\bf r})$ inside the channel. (We note
that in a steady-state situation, the external field in the vicinity to a wall
is necessary parallel to the wall.) Cummings et al. \cite{Cummings00} have
shown that for laminar incompressible flow, this boundary condition effectively
defines the flow inside the channel. Provided Eq.~(\ref{eq:mueof}) also holds
at the inlet and outlet boundaries of the channel, it becomes valid everywhere
in the channel \cite{note1}. Due to this remarkable similitude between the
steady-state velocity field of an EOF and the externally applied electric
field, the net electrophoretic velocity of nanoparticles or polyelectrolytes
can be written approximately as \begin{equation} \label{eq:mut} {\bf v}_P =
(\mueof + \mu_{e}) {\bf E}^{(ext)} =: \mu_t {\bf E}^{(ext)} \end{equation} with
the effective total mobility $\mu_t = \mueof + \mu_e$. We note that this
expression relies on two assumptions which are both not obvious: The
nanoparticles or polyelectrolytes stay well outside the diffuse layer covering
the wall, and they themselves do not influence the EOF. 

The EOF amplitude at planar walls with no-slip boundary conditions has been
calculated a long time ago by Smoluchowski \cite{Hunter89}. On the nanoscale,
however, the no-slip boundary condition does not necessarily apply. Experiments
have indicated \cite{Pit00,Tretheway02,Neto05}, that the velocity profile is
not strictly continuous at walls, {\em i.e.}, fluids exhibit a certain amoung
of slippage. This effect can be enhanced significantly by using
superhydrophobic walls which are covered by a thin gas layer.  It can also be
tuned to some extent by designing nanopatterned surfaces with alternating
hydrophobic and hydrophilic sections \cite{Barrat07}. In all of these cases,
the appropriate mesoscopic boundary condition is the 'partial slip' boundary
condition,
\begin{equation}
 \label{eq:slip}
 v_x(\pm z_B)=\mp \delta_B\frac{\partial}{\partial z} v_x(z)|_{_{z=\pm z_B}},
\end{equation}
where $z_B$ is the position of the hydrodynamic boundary (which is usually
close to the physical boundary, but not necessarily identical), and the 'slip
length' $\delta_B$ characterizes the amount of slippage.  No-slip boundaries
correspond to $\delta_B=0$, full-slip to $\delta_B \to \infty$.  From
Eq.~(\ref{eq:slip}), one would expect that the EOF amplitude is enhanced in the
presence of slippage \cite{Barrat07}, and this is indeed found in simulations
\cite{Joly04,Smiatek09}. The effect of slippage on EOF has been calculated
within the linearized Poisson-Boltzmann theory, the Debye-H\"uckel
approximation, by Joly et al. \cite{Joly04}. Below, we will derive a general 
expression which is also valid beyond the Poisson-Boltzmann theory.  

In this paper we present Dissipative Particle Dynamics (DPD)-simulations of
polyelectrolyte electrophoresis in microchannels with varying slip lengths, at
varying salt concentrations. We treat the solvent and all ions explicitly, and
all charges interact {\em via} unscreened Coulomb interactions. This allows to
investigate the interplay of solvent and polyelectrolyte, of electrostatic and
hydrodynamic interactions, of electrophoresis and EOF in full detail, with
almost no approximation.  (Our only approximation is to neglect image charge
effects, {\em i.e.}, the dielectric constant is taken to constant everywhere.)
Our results indicate that the hydrodynamic boundary conditions strongly
influence the total mobility of the polyelectrolyte, and the total mobility can
be tuned from positive to negative by varying the slip length.  The simulation
data are compared with a simple analytical expression, which is derived based
on the assumption that the flow profile in the channel follows the Stokes
equation. The numerical results are in very good agreement with the theory.

The paper is organized as follows. The theory is presented in section $2$.
Section $3$ focuses on the simulation method and the parameters used in our
simulations.  The numerical results will be shown in section $4$. We conclude
with a brief summary in section $5$.  

\section{Theoretical considerations: EOF in the presence of slippage}
\label{sec:theory}

We consider for simplicity a planar slit channel with
identical walls at $z = \pm L/2$, exposed to an external electric field $E_x$
in the $x$ direction. The electrostatic potential $\Phi(x,y,z)$ then takes the
general form $\Phi(x,y,z) = \psi(z) + E_x \: x + \mbox{const.}$ where we can
set $\psi(0)=0$ for simplicity. The electrolyte in the channel is taken to
contain $n$ different ion species $i$ with local number density $\rho_i(z)$ and
valency $Z_i$, which results in a net charge density $\rho(z) = \sum_{i=1}^n
(Z_i e) \rho_i(z)$. The electric field then generates a force density $f_x(z) =
\rho(z) E_x$ in the fluid.  Comparing the Poisson equation for the
electrostatic potential $\psi$,
\begin{equation}
 \label{eq:poisson}
 \frac{\partial^2 \psi(z)}{\partial z^2} = - \frac{\rho(z)}{\epsilon_r}
\end{equation}
(where $\epsilon_r$ is the dielectric constant), with the Stokes equation 
\cite{Lu73}
\begin{equation}
 \label{eq:stokes}
 \eta_s \frac{\partial^2 v_x(z)}{\partial z^2} = - f_x(z) = - \rho(z) E_x
 \end{equation}
(with the shear viscosity $\eta_s$), one finds immediately $\partial_{zz}
v_x(z) = \partial_{zz} \psi(z) \:(\epsilon_r \: E_x/\eta_s) $.  For symmetry
reasons, the profiles $v_x$ and $\psi$ must satisfy the boundary condition
$\partial_z v_x|_{z=0} = \partial_z \psi|_{z=0} = 0$  at the center of the
channel. This gives the relation 
\begin{equation}
 \label{eq:vx}
 v_x(z) = \frac{\epsilon_r \: E_x}{\eta_s} \psi(z) + \veof,
\end{equation}
where we have used $\psi(0)=0$ and identified the fluid velocity at the center
of the channel with the EOF velocity, $v_x(0) = \veof$. We further define
$\psi_B := \psi(\pm z_B)$ (for no-slip boundaries, $\psi_B$ is the so-called
Zeta-Potential \cite{Hunter89}). Inserting the partial-slip boundary condition
for the flow, Eq.~(\ref{eq:slip}), we finally obtain the following simple
expression for the electroosmotic mobility,
\begin{equation}
 \label{eq:mueof2}
 \mueof = {\veof}/{E_x} 
 = \mueof^0 \: (1 + \kappa \: \delta_B),
\end{equation}
where we have defined the inverse 'surface screening length'
\begin{equation}
 \label{eq:kappa}
 \kappa := \mp \frac{\partial_z \psi}{\psi} \large|_{z=\pm z_B},
\end{equation}
and $\mueof^0$ is the well-known Smoluchowski result \cite{Hunter89} for the
electroosmotic mobility at sticky walls,
\begin{equation}
 \label{eq:mueof0}
 \mueof^0 = - \epsilon_r \: \psi_B/ \eta_s.
\end{equation}

The remaining task is to determine the screening parameter $\kappa$. If the
surface charges are very small and the ions in the liquid are uncorrelated, it
can be calculated analytically within the linearized Debye-H\"uckel theory
\cite{Isr91}.  The Debye-H\"uckel equation for the evolution of the potential
$\psi$ in an electrolyte solution reads $\partial_{zz} \psi = \kappa_D^2 \psi $
with the inverse Debye-H\"uckel screening length
\begin{equation}
 \label{eq:kappa_D}
 \kappa_D = \sqrt{\frac{\sum_{i=1}^n (Z_i e)^2 \rho_{i,0}}
 {\epsilon_r \: k_{_{B}}T}},
\end{equation}
where $\rho_{i,0}$ is the density of ions $i$ far from the surface. It is
solved by an exponentially decaying function,
\begin{equation}
 \psi(z) \propto (e^{\kappa_D z} + e^{- \kappa_D z} - 2).
\end{equation}
Inserting that in Eq.~(\ref{eq:kappa}), one finds $\kappa = \kappa_D$, {\em
i.e.}, the surface screening length is identical with the Debye screening
length. Eq.~(\ref{eq:mueof0}) with $\kappa = \kappa_D$ basically corresponds to
the result of Joly et al.~\cite{Joly04}.

Unfortunately, the range of validity of the Debye-H\"uckel theory is limited,
it breaks down already for moderate surface potentials $\psi_B$ and/or for
highly concentrated ion solutions. Nevertheless, the exponential behavior often
persists even in systems where the Debye-H\"uckel approximation is not valid.
For high ion concentrations, detailed studies based on integral equations have
lead to the conclusion that the Debye-H\"uckel approximation can still be used
in a wide parameter range, if $\kappa_D$ is replaced by a modified effective
screening length \cite{Mitchell78,Kjellander94,McBride98}.  For high surface
charges, analytical solutions are again available in the so-called 'strong
coupling limit', where the profiles are predicted to decay exponentially with
the Guy-Chapman length \cite{Moreira02}. This limit is very special and rarely
encountered. At intermediate coupling regimes, the decay length must be
obtained empirically, {\em e.g.}, by fitting the charge distribution $\rho(z)$
to an exponential behavior, which is characterized by the same exponential
behavior than $\psi(z)$ by virtue of the Poisson equation, 
\begin{equation} 
  \sum_{i=q}^n (Z_i e) \: \rho_i(z) 
  \propto  \frac{\partial^2 \psi(z)}{\partial z^2} \propto
  (e^{\kappa z} + e^{- \kappa z}).
\end{equation}
Putting everything together, the total net electrophoretic mobility $\mu_t$
(Eq.~{\ref{eq:mut}) of a polyelectrolyte in the channel can be expressed in
terms of the electroosmotic mobility $\mueof$ as
\begin{equation}
\label{eq:eof_mobp}
  \frac{\mu_t}{\mueof}=1+\frac{\mu_e}{\mueof^0(1+ \kappa \: \delta_B)},
\end{equation}
where the ratio $\mu_e/\mueof^0$ depends only weakly on the ionic strength of
the electrolyte and the slip length of the surface.  The main effect of slippage
is incorporated in the factor $(1 + \kappa \: \delta_B)^{-1}$.

\section{Simulation method}

\label{sec:SM}
\subsection{Dissipative Particle Dynamics}

Newtonian fluids are effectively modeled by the Dissipative Particle Dynamics
(DPD) method \cite{Esp95,Gro97}.  DPD is a coarse-grained, momentum-conserving
simulation technique which creates a well-defined canonical ensemble.

The basic DPD equations are defined in terms of the forces on one particle, 
which involve two-particle interactions 
\begin{equation}
\label{eq:dpd}
  {\vec{F}}_{i}^{DPD}=
    \sum_{i\not={j}}{\vec{F}}_{ij}^{C}+{\vec{F}}_{ij}^{D}+{\vec{F}}_{ij}^{R}
\end{equation}
with a conservative force ${\vec{F}}_{ij}^{C}$ 
\begin{equation}
  {\vec{F}}_{ij}^{C} = -{\vec\nabla}_{ij}U_{ij}(r_{ij}).
\end{equation}
(where $r_{ij}$ denotes the distance between the centers of particles $i$ and $j$),
a dissipative force ${\vec{F}}_{ij}^{D}$ 
\begin{equation}
\label{eq:DPD}
  {\vec{F}}_{ij}^{D} =
  -\gamma_{DPD}\omega_{D}(r_{ij})(\hat{r}_{ij}\cdot{\vec{v}}_{ij})\hat{r}_{ij}
\end{equation}
with the friction coefficient $\gDPD$, and a random force $\vec{F}_{ij}^{R}$ 
\begin{equation}
\label{eq:random}
  {\vec{F}}_{ij}^{R} = \sqrt{2 \gDPD k_B T} \: 
  \omega_{R}(r_{ij})\check{\zeta}_{ij}\hat{r}_{ij}.
\end{equation}
Here $\check{\zeta}_{ij}=\check{\zeta}_{ji}$ is a symmetric random number with
zero mean and unit variance, and the weighting functions of the dissipative and
the stochastic force are related by a fluctuation-dissipation theorem,
\begin{equation}
\omega_{D}(r_{ij})=[\omega_{R}(r_{ij})]^{2}\equiv\oDPD(r_{ij}),
\end{equation}
which ensures that an equilibrium simulation samples a canonical ensemble
\cite{Esp95,Gro97}. Otherwise, the weight function $\oDPD$ is arbitrary and
will be chosen linear here as often in the literature,
 \begin{eqnarray}
      \oDPD(r_{ij})=\left \{
        \begin{array}{cc}
          1-\frac{r_{ij}}{r_c} & :  r_{ij} < r_c \\
          0 & :  r_{ij} \geq r_c
        \end{array}
      \right.
    \end{eqnarray}
where $r_c$ denotes the cut-off radius.

\subsection{Tunable slip boundaries}

The hydrodynamic boundary condition at the walls is realized with a recently
developed method \cite{Smiatek08} that allows to implement arbitrary
partial-slip boundary conditions: We introduce an additional
coordinate-dependent viscous force that mimicks the wall/fluid friction
\begin{equation}
\label{eq:langevin}
  {\bf{F}}_i^{L}={\bf{F}}_i^{D}+{\bf{F}}_i^{R}
\end{equation}
with a dissipative contribution
\begin{equation}
  {\bf{F}}_i^{D}=-\gamma_L\: \omega_L(z) \: \: ({\bf{v}}_i-{\bf{v}}_{wall})
\end{equation}
coupling to the relative velocity $({\bf{v}}_i-{\bf{v}}_{wall})$ of the particle
with respect to the wall, and a stochastic force
\begin{equation}
  F_{i,\alpha}^R= \sqrt{2 \gamma_L \: k_B T \: \omega_L(z)}\;\chi_{i,\alpha}
\end{equation}
which satisfies the fluctuation-dissipation relation and thus ensures that the
local equilibrium distribution is again a Boltzmann distribution.  Here
$\alpha$ is $\alpha = x,y,z$ and $\chi_{i,\alpha}$ is a Gaussian distributed
random variable with mean zero and variance one: $\langle \chi_{i,\alpha}
\rangle = 0$, $\langle \chi_{i,\alpha} \chi_{j,\beta} \rangle = \delta_{ij}
\delta_{\alpha \beta}$.  The viscous coupling between fluid and wall is
achieved by the locally varying viscosity $\gamma_L \omega_L(z)$ with
$\omega_L(z) = 1 - z/z_c$ up to a cut-off distance $z_c$.  The prefactor
$\gamma_L$ can be used to tune the strength of the friction force and hence the
value of the slip length.  Within this approach it is possible to tune the slip
length $\delta_B$ systematically from full-slip to no-slip, and to derive an
analytic expression for the slip length as a function of the model parameters
\cite{Smiatek08}.

\subsection{Simulation details}
\label{sec:SD}

We have studied the electrophoresis of charged polymers of length $N=20$ in
electrolyte solutions, confined by a planar slit channel with charged walls.
All particles, polymer, solvent and ions, are modeled explicitly. We use a
simulation box of size ($12\sigma\times 12\sigma\times 10\sigma$) which is
periodic in $x$- and $y$-direction and confined by impermeable walls in the
$z$-direction. The walls repel the particles {\em via} a soft repulsive WCA
potential \cite{WCA} of range $\sigma$ and amplitude $\epsilon$. (Hence the
accessible channel width for the particles is actually $L_z = 8 \sigma$).  Ions
and monomers repel each other with the same WCA potential. In addition, chain
monomers are connected by harmonic springs 
\begin{equation}
  U_{harmonic}=\frac{1}{2}{k}(r_{ij}-r_0)^2
\end{equation}
with the spring constant $k=25 \epsilon/\sigma^2$ and $r_0=1.0\sigma$.  Neutral
solvent particles have no conservative interactions except with the walls. 

The wall contains immobilized, negatively charged particles at random
positions.  Every second monomer on the polyelectrolyte carries a negative
charge. The solvent contains the positive counterions for the walls and the
polyelectrolyte, and additional (positive and negative) salt ions. All charges
are monovalent, and the system as a whole is electroneutral. In addition to
their other interactions, charged particles interact {\em via} a Coulomb
potential with the Bjerrum length $\lambda_B=e^2/4\pi\epsilon_r
k_BT=1.0\sigma$, and they are exposed to an external field $E_x=-1.0
\epsilon/e\sigma$. Specifically, we have studied systems with a surface charge
density of $\sigma_A=-0.208e\sigma^{-2}$. The total counterion density was
$\rho =0.06\sigma^{-3}$ and the salt density varied between
$\rho_s=$0.05625, 0.0375, 0.03, 0.025, and $0.015 \sigma^{-3}$.

We use DPD simulations with a friction coefficient
$\gDPD=5.0\sigma^{-1}(m\epsilon)^{1/2}$.  The density of the solvent particles
was $\rho=3.75\sigma^{-3}$, and the temperature of the system was
$T=1.0\epsilon/k_B$. For these parameters, the shear viscosity of the DPD fluid
-- as determined by fitting the amplitude of Plane Poiseuille flows
\cite{Smiatek08} -- is given by
$\eta_s=(1.334\pm0.003)\sigma^{-2}(m\epsilon)^{1/2}$.  The DPD timestep was
$\delta t=0.01\sigma(m/\epsilon)^{1/2}$. 

Tunable-slip boundary conditions were used with friction coefficients
$\gamma_L=$0.1, 0.25, 0.5, 0.75, 1.0, and $6.1 \sigma^{-1}(m\epsilon)^{1/2}$.
The range of the viscous layer was $z_c=2.0\sigma$.  Only the solvent particles
interact with the tunable-slip boundaries.  By performing Plane Poiseuille and
Plane Couette flow simulations with the above given parameters, the slip length
$\delta_B$ and the hydrodynamic boundary positions $z_B$ can be determined
independently \cite{Smiatek08}.  The hydrodynamic boundary position is found at
$|z_B|=(3.866\pm 0.266)\sigma$ in all simulations. The corresponding slip
lengths are presented in Table~\ref{tab:1} together with the theoretical values
predicted by the analytic expression  in Ref. $38$.  The comparison
shows that the simulated results are in good agreement with the theory.

\begin{table}
\caption{Slip lengths $\delta_B$ for different layer friction coefficients 
  $\gamma_L$, compared with  theoretical value $\delta_B^T$ according to
  Ref. $38$.}
\label{tab:1}       
\begin{tabular}{llll}
\hline\noalign{\smallskip}
$\gamma_L[\sigma^{-1}(m\epsilon)^{1/2}]$ & $\delta_B [\sigma]$ & $\pm
\delta_B [\sigma]$ & $\delta_B^T [\sigma]$\\
\noalign{\smallskip}\hline\noalign{\smallskip}
$0.1$ & $14.977$ & $1.879$ & $14.000$\\
\noalign{\smallskip}\hline\noalign{\smallskip}
$0.25$ & $5.664$ & $0.783$ & $5.458$\\
\noalign{\smallskip}\hline\noalign{\smallskip}
$0.5$ & $2.626$ & $0.521$ & $2.613$\\
\noalign{\smallskip}\hline\noalign{\smallskip}
$0.75$ & $1.765$ & $0.409$ & $1.664$ \\
\noalign{\smallskip}\hline\noalign{\smallskip}
$1.0$ & $1.292$ & $0.423$ & $1.190$\\
\noalign{\smallskip}\hline\noalign{\smallskip}
$6.1$ & $0.000$ & $0.197$ & $0.000$\\
\noalign{\smallskip}\hline
\end{tabular}
\end{table}

The electrostatics were calculated by P3M \cite{Hockney81} and the ELC
(electrostatic layer correction)-algorithm \cite{Arnold02} for $2D+h$ slabwise
geometries.  All simulations have been carried out with the freely available
software package {\sf ESPResSo} \cite{Espresso1,Espresso2}

\section{Numerical results}
\label{sec:nr}

\begin{figure}[tbh]
  \includegraphics[width=0.4\textwidth]{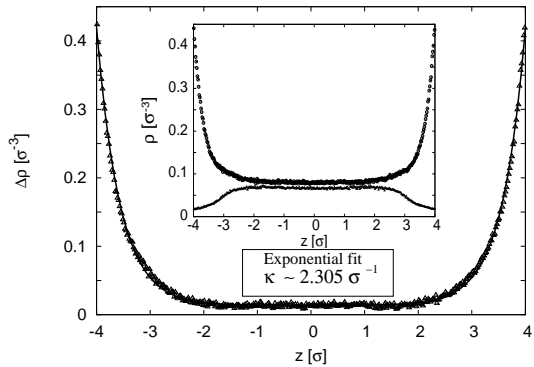}
\caption{Distribution of the ionic difference $\Delta \rho=\rho_c-\rho_a$ 
  between cations and anions in the solution (not counting the polyelectrolyte)
  for an exemplary salt concentration of $\rho_s= 0.05625\sigma^{-3}$ and the
  surface charge density $\sigma_A=-0.208e\sigma^{-2}$. The black line corresponds
  to an exponential fit (Eq.~(\protect\ref{eq:fit})) with an effective inverse 
  screening length of $\kappa=2.305\pm0.025\sigma^{-1}$.
  {\bf Inset:} Distribution of cations (circles) and anions (triangles)
  for the same system.} 
\label{fig:ionscomb}       
\end{figure} 

Fig.~\ref{fig:ionscomb} shows the average ionic distributions of anions $\rho_a$
and cations $\rho_c$ for the salt concentration $\rho_s=0.05625\sigma^{-3}$.  
Here the 'cations' include the positively charged salt ions and the counterions 
of the wall and the polyelectrolyte, and the 'anions' only the negatively charged 
salt ions. Due to the presence of the polyelectrolyte in the middle of the channel, 
the average cation density there is slightly increased. To determine the inverse
effective screening length $\kappa$, we have thus fitted the following function
\begin{equation}
\label{eq:fit}
  \Delta \rho = \Delta\rho_0(e^{-\kappa z}+e^{\kappa z})+c
\end{equation}

to the ionic difference $\Delta \rho = \rho_c-\rho_a$. The exponential fit
describes the data very well (black solid line in Fig.~\ref{fig:ionscomb}).
The fit parameters for $\kappa$ are listed in Table \ref{tab:2}, along with the
values for the Debye-H\"uckel screening parameter $\kappa_D$
(Eq.~(\ref{eq:kappa_D})). The decay lengths are overall very different from
those predicted by the Debye-H\"uckel theory.  We conclude that the system is
outside the validity region of the linearized Poisson-Boltzmann approximation.
This is perhaps not surprising, given that the individual ion profiles (inset
of Fig.~\ref{fig:ionscomb}) at the walls deviate strongly from their bulk
value, {\em i.e.}, these deviations can hardly be considered as small
perturbations.  The surface charge is too high. On the other hand, the
electrostatic coupling constant $\Xi = 2 \pi Z^3 \lambda_B^3 \sigma_A  \sim
0.2$ ($Z=1$ is the valency of the cations) is still much smaller than unity,
hence we are still in a 'weak coupling' regime. This is also evident from the
fact that the effective screening parameter $\kappa$ differs strongly from the
Guy Chapman length, $\mu^{-1} = 2 \pi \lambda_B Z \sigma_A = 1.31 \sigma$.

\begin{table}[b]
\caption{Fitted inverse screening lengths $\kappa$ and Debye-H\"uckel screening
  parameter $\kappa_D$ for different salt concentrations $\rho_s$ and the fixed 
  counterion density of $\rho=0.06\sigma^{-3}$.}
\label{tab:2}       
\begin{tabular}{llll}
\hline\noalign{\smallskip}
$\rho_s[\sigma^{-3}]$ & $\kappa [\sigma^{-1}]$ & $\pm\kappa[\sigma^{-1}]$
 & $\kappa_D [\sigma^{-1}]$ \\
\noalign{\smallskip}\hline\noalign{\smallskip}
0.015 & 1.996 & 0.041 & 0.98 \\
\noalign{\smallskip}\hline\noalign{\smallskip}
0.0225 & 2.011 & 0.049 & 1.02\\
\noalign{\smallskip}\hline\noalign{\smallskip}
0.03 & 1.983 & 0.041 & 1.07 \\
\noalign{\smallskip}\hline\noalign{\smallskip}
0.0375 & 2.182 & 0.047 & 1.11\\
\noalign{\smallskip}\hline\noalign{\smallskip}
0.05625 & 2.305 & 0.025 & 1.21\\
\noalign{\smallskip}\hline
\end{tabular}
\end{table}

\begin{figure}[tbh]
  \includegraphics[width=0.4\textwidth]{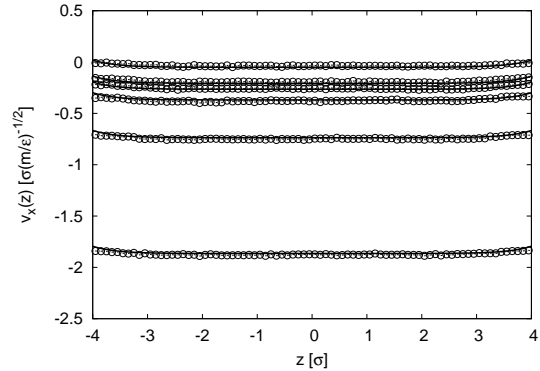}
  \caption{Exemplary flow profiles for a salt concentration 
    $\rho_s=0.05625\sigma^{-3}$ for varying slip lengths (from bottom to top:
    $\delta_B=(14.98, 5.66, 2.63, 1.77, 1.29, 0.00) \sigma$.) The black lines are
    the theoretical predictions obtained by integrating the Stokes equation
    (Eq.~(\ref{eq:stokes})) with a fitted inverse screening
    length of $\kappa=2.305 \sigma^{-1}$.}
\label{fig:solvflow}       
\end{figure}

\begin{figure}[tbh]
  \includegraphics[width=0.4\textwidth]{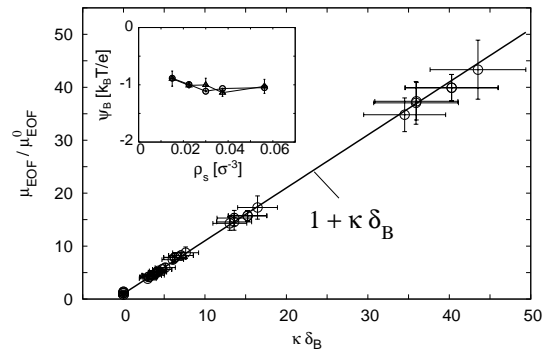}
\caption{Ratio $\mueof/ \mueof^0$ plotted against $\delta_B\kappa$
  for the different salt concentrations and screening lengths given in
  Table~\ref{tab:1} and \ref{tab:2}. The black line is the theoretical 
  prediction of Eq.~(\ref{eq:mueof2}) with slope $1+\kappa \: \delta_B$. 
  {\bf Inset:} Surface potential as obtained from $\mueof^0$ using 
  Eq.~(\ref{eq:mueof0}) (circles) and indepently by a test charge method
  (triangles) as a function of the salt concentration $\rho_s$.}
\label{fig:solvcscale}       
\end{figure}

The EOF profiles for the same salt concentration ($\rho_s=0.05625\sigma^{-3}$)
are shown in Fig.~\ref{fig:solvflow}. The different curves correspond to
different hydrodynamic boundary conditions (slip lengths). As expected, the
flow velocity increases drastically for larger slip lengths. All curves are in
good agreement with the theoretical predictions, which were obtained by
integrating the Stokes equation (\ref{eq:fit}) numerically with the correct
partial-slip boundary conditions.  In agreement with our earlier studies at
zero salt concentration \cite{Smiatek09}, we thus find that a description based
on the Stokes equation -- a continuum equation -- remains valid even for very
narrow channels.

The flow velocity in the middle of the channel gives the EOF mobility.
Fig.~\ref{fig:solvcscale} compares our numerical results for all salt
concentrations and slip lengths with the theoretical prediction of
Eq.~(\ref{eq:mueof2}), where $\mueof^0$ has been determined by a linear
regression for each salt concentration independently. We find good agreement
between simulation data and theory. This confirms the validity of our
theoretical result, Eq.~(\ref{eq:mueof2}). It also demonstrates that the
polyelectrolyte, which was present in all simulations, does not perturb the EOF
even in very narrow channels.

The values of the EOF mobility for zero slip length, $\mueof^0$, can be used to
determine the surface potential $\psi_B$ for the different salt concentrations {\em
via} Eq.~(\ref{eq:mueof0}). As a consistency check, we have also determined
$\psi_B$ independently by inserting a test charge into the ion layer at $z =
z_B$.  The results are shown in the inset of Fig.~\ref{fig:solvcscale}. Both
methods give identical results. The surface potential is found to be largely
independent of the salt concentration.

\begin{figure}[t]
  \includegraphics[width=0.4\textwidth]{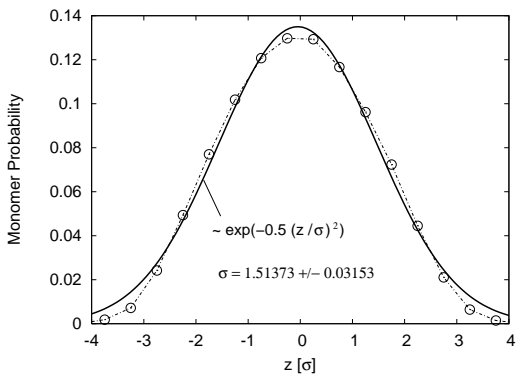}
\caption{Normalized monomer distribution inside the channel for the salt
  concentration $\rho_s=0.05625\sigma^{-3}$.
} 
\label{fig:mono20dens}       
\end{figure}

After investigating the EOF of the solvent, we discuss the properties of the
polyelectrolyte. The probability distribution for finding a monomer at a given
position $z$ is shown in Fig.~\ref{fig:mono20dens} for the salt concentration
$\rho_s=0.05625\sigma^{-3}$. It is approximately Gaussian with a peak in the
middle of the channel and a variance $\mbox{Var} \sim 2.28\sigma$. Thus the
polyelectrolyte mainly 'senses' the EOF in the middle of the channel, and the
details of the flow profiles close to the channel walls have very little
influence on its net mobility: The assumption that the total mobility 
is governed by a single EOF velocity $\vveof$ (Eq.~(\ref{eq:mut})) is probably 
legitimate.

\begin{table}[b]
\caption{Radius of gyration $R_g$ and end to end radius $R_e$ for a
  polyelectrolyte with $N=20$ monomers for different salt concentrations $\rho_s$.}
\label{tab:3}       
\begin{tabular}{lll}
\hline\noalign{\smallskip}
$\rho_s[\sigma^{-3}]$ & $R_g [\sigma]$ & $R_e[\sigma]$ \\
\noalign{\smallskip}\hline\noalign{\smallskip}
0.015 & $3.2218\pm 0.047$  & $10.6480\pm0.0314$ \\
\noalign{\smallskip}\hline\noalign{\smallskip}
0.0225 & $3.1661\pm0.0041$ & $10.2736\pm0.0266$ \\
\noalign{\smallskip}\hline\noalign{\smallskip}
0.03 & $3.1486\pm0.0451$ & $10.1777\pm0.0292$ \\
\noalign{\smallskip}\hline\noalign{\smallskip}
0.0375 & $3.1279\pm0.0045$ & $10.0819\pm0.0287$ \\
\noalign{\smallskip}\hline\noalign{\smallskip}
0.05625 & $3.0825\pm0.0045$ & $9.8331\pm0.0280$ \\
\noalign{\smallskip}\hline
\end{tabular}
\end{table}  

The influence of the ion profiles on the chain structure of the polyelectrolyte
can be investigated by considering static properties like the radius of
gyration $R_g^2=(1/2N^2)\sum_{i,j=1}^N<(\vec{R}_i-\vec{R}_j)^2>$ and the
end-to-end radius $R_e^2 = <(\vec{R}_N-\vec{R}_1)^2>$ \cite{Doi86}.  The
results for these parameters are shown in Table~\ref{tab:3}. Both
characteristic lengths decrease with increasing salt concentration due to a
more effective screening of electrostatic interactions, in accordance with
standard theories \cite{Dob05}. The ratio between the end-to-end radii and the
gyration radii is unusually large, which is most likely a squeezing effect due
to the presence of the channel walls \cite{deGennesbook,Canna08}. Specific
flow-induced effects such as shear-induced elongation \cite{Boro05} are
probably less important, since the flow profile is basically constant inside
the channel. It should be noted that the pure electrophoretic mobility $\mu_e$
of the chain is presumably modified by the confinement as an indirect effect of
the elongation. This effect has not been investigated in the present study.

\begin{figure}[t]
  \includegraphics[width=0.4\textwidth]{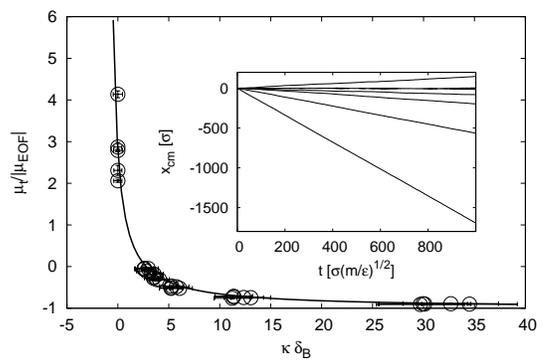}
\caption{Ratio $\mu_t/|\mueof|$ plotted against $\delta_B \: \kappa$ for all
  slip lengths (Table \ref{tab:1}) and salt concentrations (Table \ref{tab:2}).
  The black line is the theoretical prediction of Eqn.~(\ref{eq:eof_mobp}) with
  absolut values of $|\mueof|$. In the limit 
  $\delta_B \: \kappa\rightarrow \infty$, the total mobility of the
  polyelectrolyte is equal to the electroosmotic mobility $\mueof$. The ratio
  $\mu_e/\mueof^0$ has been fitted to $-3.778\pm 0.128$. Negative values
  of $\mu_t/|\mueof|$ indicate negative total mobilities of the polyelectrolyte.
  {\bf Inset:}  
  Total displacement of the polyelectrolytes center of mass for different
  boundary conditions and a salt concentration of
  $\rho_s=0.05625\sigma^{-3}$. The total mobility becomes negative for
  $|\mu_e|\ll|\mueof|$. The lines correspond from top to bottom
  to the slip lengths $\delta_B\approx (0.00, 1.292, 1.765, 2.626, 5.664,
  14.98)\sigma$. Thus larger slip lengths indirectly enhance the total 
  mobility of the polyelectrolyte.}   
\label{fig:mobscalefree}       
\end{figure}

The total mobility of the polyelectrolyte for varying boundary conditions and
salt concentrations is finally presented in Fig.~\ref{fig:mobscalefree}.  The
theoretical prediction of Eq.~(\ref{eq:eof_mobp}) agrees well with the
numerical results with the single fit parameter $\mu_e/\mueof^0=-3.778\pm
0.128$.  It is remarkable that this parameter can be set to a constant, {\em
i.e.}, it seems to be largely independent of the salt concentration $\rho_s$.
Since $\mueof^0$ does not depend on $\rho_s$ (see Fig.~\ref{fig:solvcscale},
inset), this means that $\mu_e$ is also independent of $\rho_s$ for the range
of salt concentrations considered here \cite{Muthu97}.

For no-slip boundary conditions with $\delta_B\approx 0$, we find ordinary 
behaviour where the polyelectrolyte follows the applied electric field.
In the presence of wall slip, however, the EOF becomes stronger and
eventually dominates. Then the total mobility may become negative,
{\em i.e.}, the polyelectrolyte migrates in a direction which is opposite
to the applied force. The inset of Fig.~\ref{fig:mobscalefree} 
illustrates this by showing the total displacement of the chain's 
center of mass for the salt concentration $\rho_s=0.05625\sigma^{-3}$ and 
various slip lengths.  In nearly all cases except $\delta_B\approx 0$, the 
total mobility of the polyelectrolyte is negative.

To summarize this section, both the assumptions and the predictions of section
\ref{sec:theory} are supported by our numerical results.  The total mobility of
the polyelectrolyte can therefore be adequately described by
Eqs.~(\ref{eq:mueof2}) and (\ref{eq:eof_mobp}).

\section{Conclusions}

We have presented mesoscopic DPD simulations of polyelectrolyte electrophoresis
in narrow microchannels, taking full account of hydrodynamic and electrostatic
interactions. A particular focus was put on studying the effects of the
hydrodynamic boundary conditions at the channel walls on the electroosmotic
flow and on the net electrophoretic mobility of the polyelectrolyte. We have
shown that they can be incorporated into a single dimensionless parameter
$(1+\kappa \: \delta_B)$, where $\delta_B$ is the slip length and $\kappa$ the
(local) inverse screening length of the charge distribution at the wall. This
was derived analytically and supported by our numerical data. It remained valid
even for very narrow channels, where the chain conformations were affected by
the confinement.

We have shown that wall slip massively enhances the EOF and hence
influences the total mobility of the polyelectrolyte. If the EOF mobility 
$\mueof$ and the free draining mobility $\mu_e$ oppose each other, 
{\em i.e.}, if the effective charges on the polyelectrolyte and the walls 
have the same sign, the mobility may even become negative.  As mentioned 
in the introduction, this effect has also been observed experimentally 
\cite{Mathe07}. In the other case, where the sign of the charges on the 
polyelectrolyte and the wall are opposite, the main effect of slip is
to enhance the total mobility of the polyelectrolyte.

In summary, the total mobility of polyelectrolytes in microchannels
results from an interplay of electroosmotic, electrophoretic,
electrostatic and slippage effects. The latter have a particularly
strong influence and can be used to design channels with improved properties. 
For example, the characteristics of the channel walls could be designed
to tune effective slip lengths \cite{Barrat07} and hence flow velocities, 
which offers the possibility to optimize the time which is needed for polymer 
migration or separation techniques. This could be an important aspect for 
future applications in microchannels or micropumps to accelerate measuring times.

\begin{acknowledgements}
We thank Christian Holm, Burkhard D{\"u}nweg, Ulf D. Schiller, Marcello Sega
and Kai Grass for nice and fruitful discussions and the Arminius PC$^2$
Cluster at Paderborn University, HLRS Stuttgart and NIC
J{\"u}lich for computing time. 
J.~S.~ especially thanks Stefanie G{\"u}rtler and Theodor A. Smiatek.
Financial funding from the Volkswagen Stiftung is gratefully acknowledged.
\end{acknowledgements}



\end{document}